\documentclass{PoS}

\usepackage{microtype}

\title{Nuclear PDF constraints from p+Pb collisions at the LHC}

\ShortTitle{Nuclear PDF constraints from p+Pb collisions at the LHC}

\author{\speaker{Ilkka Helenius}$^{\mbox{}\,a}$, Hannu Paukkunen$^{bc}$ and Kari J.~Eskola$^{bc}$\\
\llap{$^a$}Department of Astronomy and Theoretical Physics, Lund University, S\"{o}lvegatan 14A,\\ SE-223 62 Lund, Sweden\\
\llap{$^b$}University of Jyvaskyla, Department of Physics, P.O. Box 35,\\ FI-40014 University of Jyvaskyla, Finland\\
\llap{$^c$}Helsinki Institute of Physics, P.O. Box 64, FI-00014 University of Helsinki, Finland\\
        E-mail: \email{ilkka.helenius@thep.lu.se}, \email{hannu.t.paukkunen@jyu.fi}, \email{kari.eskola@jyu.fi}}

\abstract{As the current nuclear PDF analyses are mainly constrained by fixed-target Drell-Yan and deeply inelastic scattering data only the quark nuclear modifications at fairly large $x$ values are in a good control. Inclusive pion production in d+Au collisions at RHIC provides some constraints for gluons but due to the limited kinematic reach of the data the gluon modifications remain uncertain especially at small values of $x$. In this talk, we discuss how the existing data from p+Pb collisions at the LHC can improve the nuclear PDF fits and which measurements would be sensitive to the small-$x$ gluons. In particular, we consider inclusive hadron production, compare this to direct photons, and show estimates of the effect of CMS dijet measurements to the uncertainty of nuclear gluon distributions.}

\FullConference{XXIII International Workshop on Deep-Inelastic Scattering\\
		27 April - May 1 2015\\
		Dallas, Texas}

\begin{document}

\section{Introduction}

To calculate cross sections in any hadronic collisions the structure of the colliding hadrons must be known. This structure cannot yet be calculated directly from QCD but using experimental data and the DGLAP evolution equations the parton distribution functions (PDFs) describing the hadron structure within collinear factorization can be extracted. Thus, the precision of PDFs depends directly on the data used in the analysis. For the free proton analyses there are plenty of precise data available and nowadays different analysis groups provide very accurate global fits that are in a good agreement with each other \cite{Rojo:2015acz}. For the nuclear PDFs (nPDFs) the situation is not as good: The current fits \cite{Eskola:2009uj, Hirai:2007sx, deFlorian:2011fp, Kovarik:2015cma} rely mostly on the fixed-target deep inelastic scattering (DIS) and Drell-Yan (DY) data where the kinematic reach is limited to rather high values of $x$ and that are mainly sensitive to quark nPDFs. The gluons are constrained indirectly via DGLAP evolution and some analyses use also inclusive pion production data from d+Au collisions at RHIC to have more direct information on the gluon modifications. The kinematic reach of the pion production data is, however, rather limited and different analysis groups even have mutually contradictory interpretations for the data \cite{Eskola:2012rg} which underscores the need for further constraints. Here, we study how the recent measurements in p+Pb collisions at the LHC can improve the fits and discuss which LHC measurements would help to further constrain nPDFs.

\section{Inclusive hadron production}

\begin{figure}[htb]
\begin{minipage}[t]{0.49\textwidth}
\centering
\includegraphics[width=\textwidth]{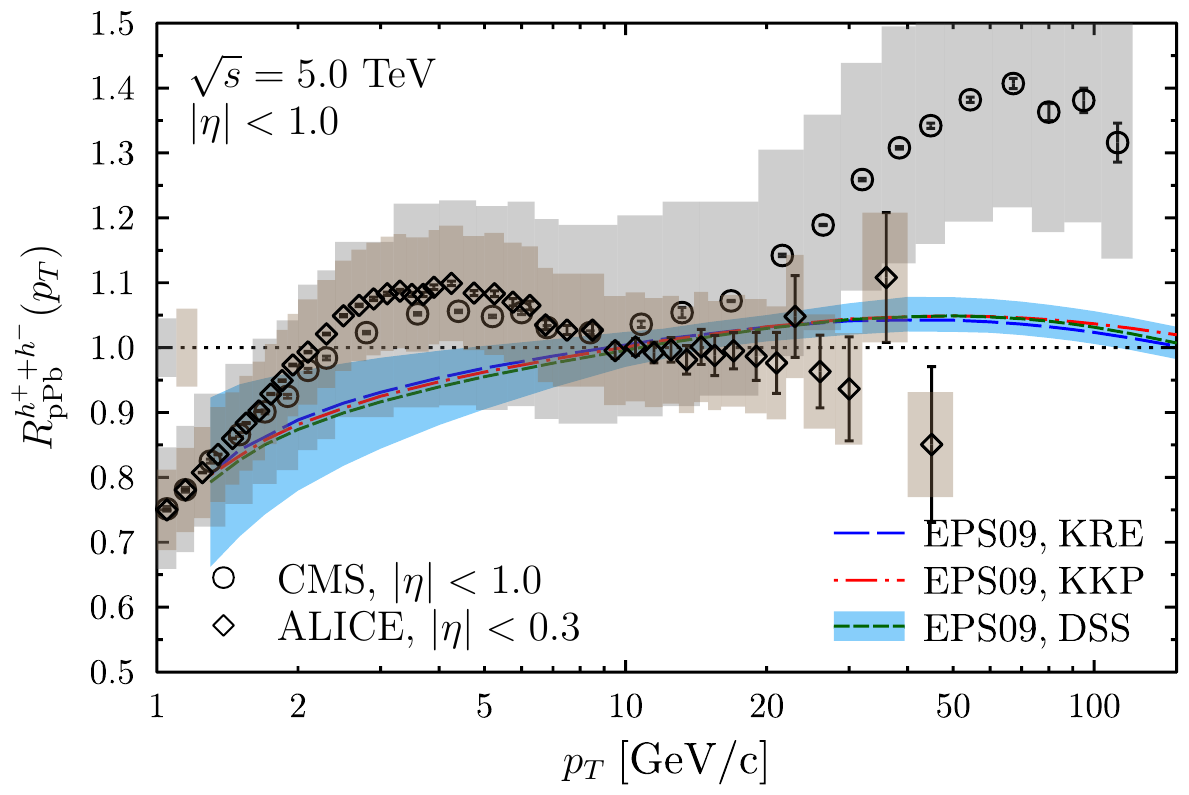}
\caption{Nuclear modification factor for inclusive charged hadrons in p+Pb collisions at $\sqrt{s_{\rm_{NN}}}=5.0\,\mathrm{TeV}$. Data are from ALICE \cite{Abelev:2014dsa} (diamonds) and CMS \cite{Khachatryan:2015xaa} (circles) and the NLO pQCD results are calculated using CT10 PDFs with EPS09 nuclear modification and three FFs, see Ref.~\cite{Helenius:2012wd} for details.}
\label{fig:RpPb_ch}
\end{minipage}%
\hspace{0.02\linewidth}
\begin{minipage}[t]{0.49\textwidth}
\centering
\includegraphics[width=\textwidth]{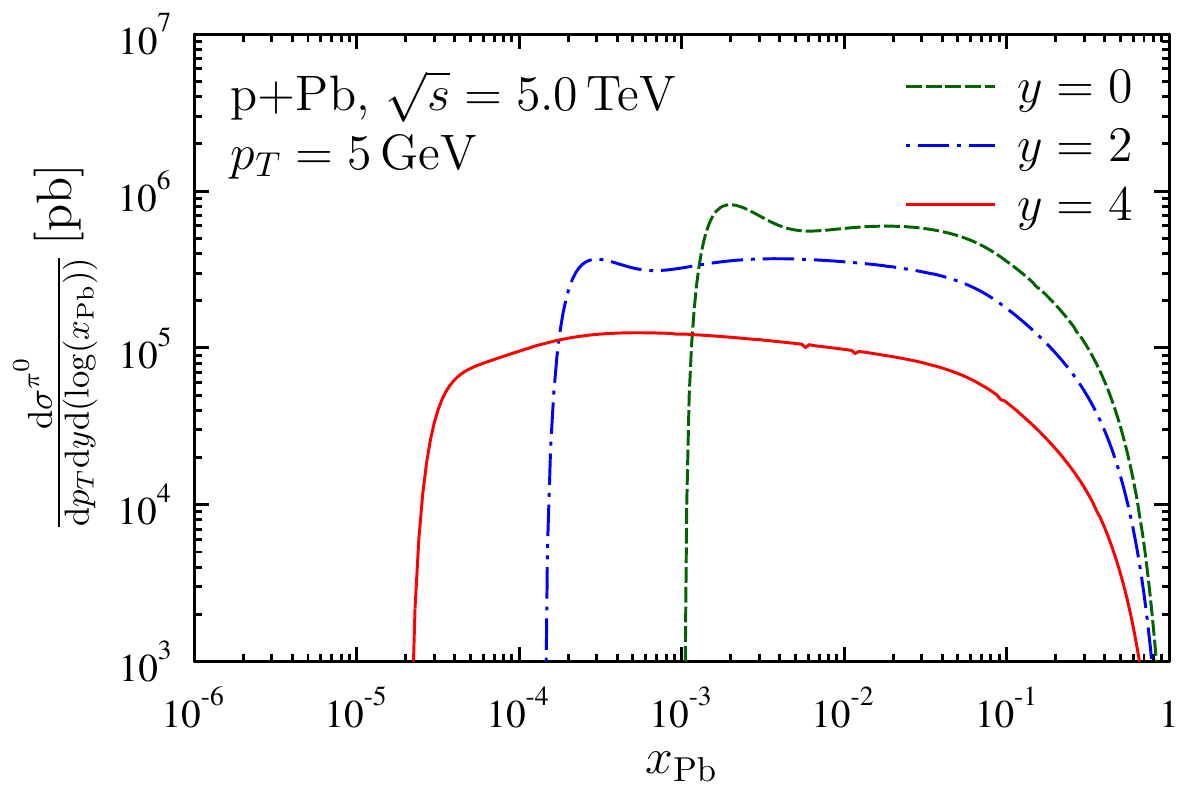}
\caption{The contribution to differential cross section from different values of $x_{\rm Pb}$ for inclusive $\pi^0$ production in p+Pb collisions at $\sqrt{s_\mathrm{_{NN}}}=5.0\,\mathrm{TeV}$ and $p_{\rm T} = 5\,\mathrm{GeV/c}$ for $\eta = 0$ (dashed green), $\eta = 2$ (dot-dashed blue) and $\eta = 4$ (solid red). From Ref.~\cite{Helenius:2014qla}.}
\label{fig:dsigmadx2pi0}
\end{minipage}
\end{figure}

The nuclear effects in p+Pb collisions at the LHC can be conveniently studied by defining the nuclear modification factor $R_{\rm pPb}$ as a ratio between the cross section in minimum bias 
proton-lead collisions and the corresponding cross section in proton-proton collisions,
\begin{equation}
R_{\rm pPb} = \frac{1}{208}\frac{\mathrm{d}^2\sigma_{\rm pPb}}{\mathrm{d}p_T\mathrm{d}\eta}\bigg/ \frac{\mathrm{d}^2\sigma_{\rm pp}}{\mathrm{d}p_T\mathrm{d}\eta}.
\end{equation}
Three LHC experiments (ALICE \cite{Abelev:2014dsa}, CMS \cite{Khachatryan:2015xaa} and ATLAS \cite{ATLAS:2014cza}) have published data for inclusive charged hadron production at mid-rapidity in p+Pb collisions. A comparison between the EPS09-based prediction \cite{Helenius:2012wd} and the ALICE and CMS measurements is shown in figure \ref{fig:RpPb_ch}. At low $p_{\rm T}$ (around $p_{\rm T}\sim 3\,\mathrm{GeV/c}$) the data overshoots the prediction by some 20\%. This was not completely unexpected as the baryon production is not very well described with leading-twist pQCD at that region and a comparison with identified mesons would be preferred. Indeed, the preliminary ALICE result for charged pions \cite{Knichel:2014yaa} does not show a similar enhancement at few GeV/c. What is more surprising is the distinct rise of the CMS data at $p_{\rm T}\gtrsim 20\,\mathrm{GeV/c}$. The observed 40 \% enhancement at high-$p_{\rm T}$ is way above any nPDF-based prediction and needs to be carefully studied before the data can be included into the nPDF fits. Similar trend is seen also in the preliminary ATLAS data but as they have used some cuts on centrality we do not include this data to the comparison here.

The calculation of inclusive hadron spectra involves a convolution with fragmentation functions (FFs). For this, even the double-differential cross sections receive contributions from a broad range of $x$. This is illustrated in figure \ref{fig:dsigmadx2pi0} showing the contribution to differential cross section at $p_{\rm T} = 5\,\mathrm{GeV/c}$ and $\eta=0,2,4$ from different values of $x_{\rm Pb}$ in the case of inclusive pion production at $\sqrt{s_\mathrm{_{NN}}}=5.0\,\mathrm{TeV}$. The rapidities are defined in the nucleon-nucleon CMS-frame and the positive values correspond to proton-going direction. Typically the suppression (shadowing) turns to an enhancement (antishadowing) around $x\sim 0.01$ in the nPDF fits. As the cross section at mid-rapidity is sensitive to both of these regions these competing effects partly cancel each other out. This cancellation means that it is not possible to measure these effects unambiguously using $R_{\rm pPb}^{h^++h^-}$ alone but some other complementary measurements are required.

\section{Dijets}

\begin{figure}[htb]
\begin{center}
\includegraphics[width=\textwidth]{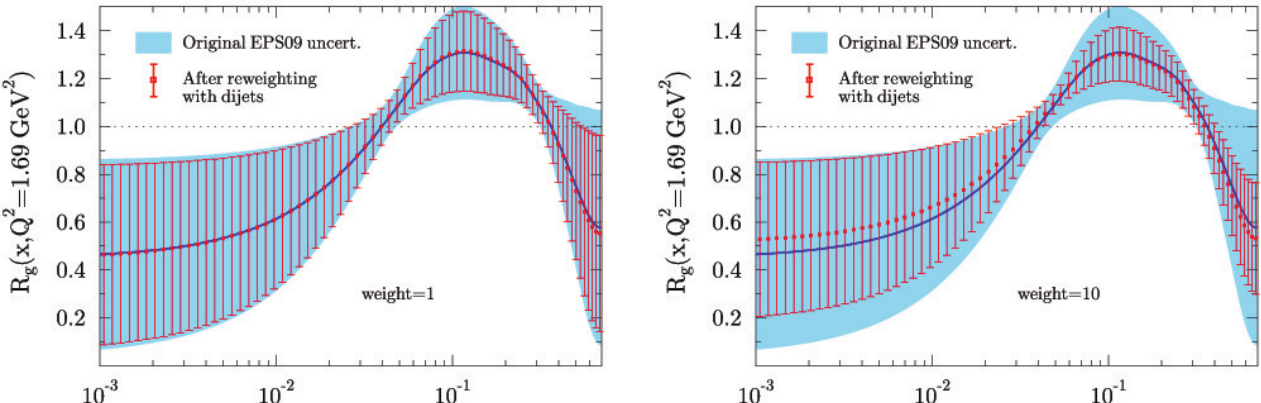}
\caption{The gluon nuclear modification for Pb nucleus and its uncertainty from the original EPS09 analysis (blue band) and after reweighting with the dijet data (red bars) using a unit weight (left panel) and a weight factor 10 (right panel). From Ref.~\cite{Paukkunen:2014pha}.}
\label{fig:dijetReWeight}
\end{center}
\end{figure}

A very promising new observable to constrain the gluons in the antishadowing and EMC regions is the dijet production in p+Pb collisions. The first data recently published by CMS \cite{Chatrchyan:2014hqa} clearly support an EPS09-style gluon antishadowing and EMC-effect. To study the potential effect of these data on nPDFs a Hessian reweighting analysis has been carried out in Ref.~\cite{Paukkunen:2014pha}. The results of this analysis for the gluon modification are reproduced here in figure \ref{fig:dijetReWeight}. If no weight is applied for the dijet data the impact on the fit precision stays more or less the same but when applying a weight factor of 10 the estimated uncertainty of the fit is clearly reduced. However, the latter result should actually be closer to the ``true'' impact of these data recalling that the importance of the RHIC pion data was increased in the EPS09 analysis by a factor of 20. This signals that the dijet measurements will provide important constraints for the mid/large-$x$ behavior of the gluon modification in future global analyses.

\section{Direct photons}

\begin{figure}[htb]
\begin{center}
\includegraphics[width=0.49\textwidth]{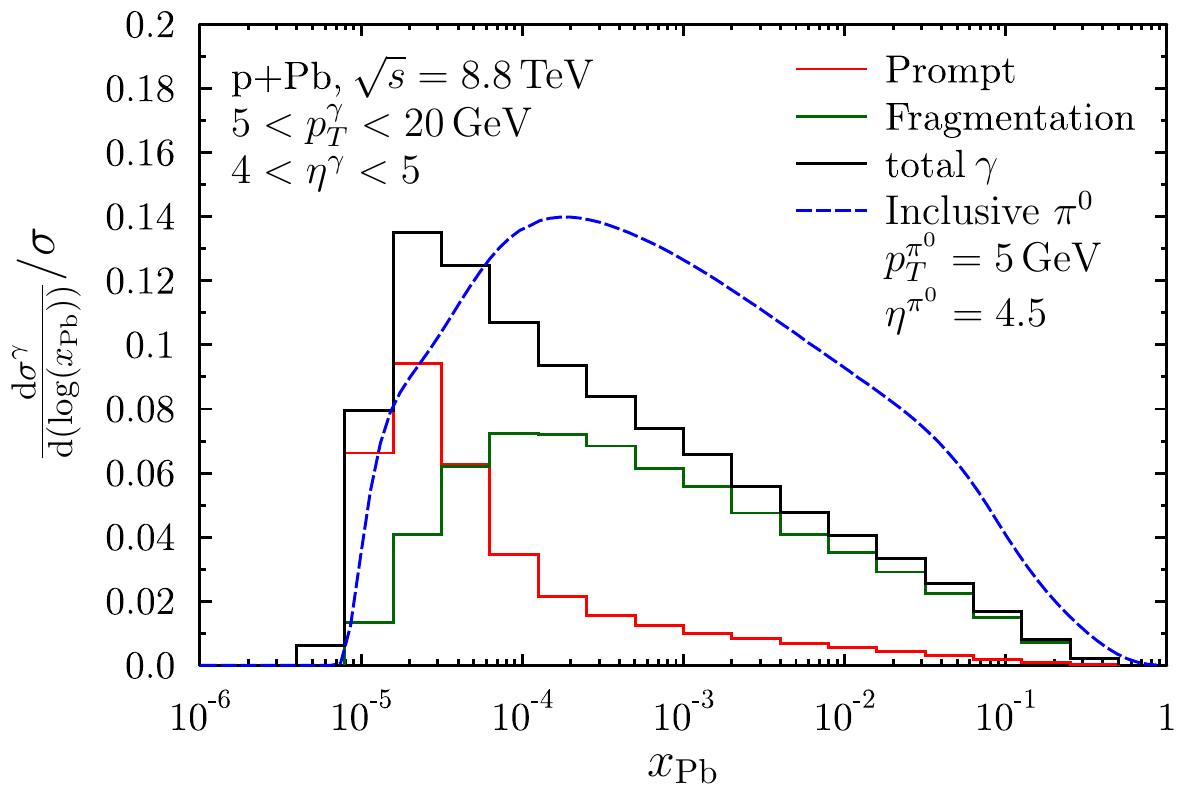}
\includegraphics[width=0.49\textwidth]{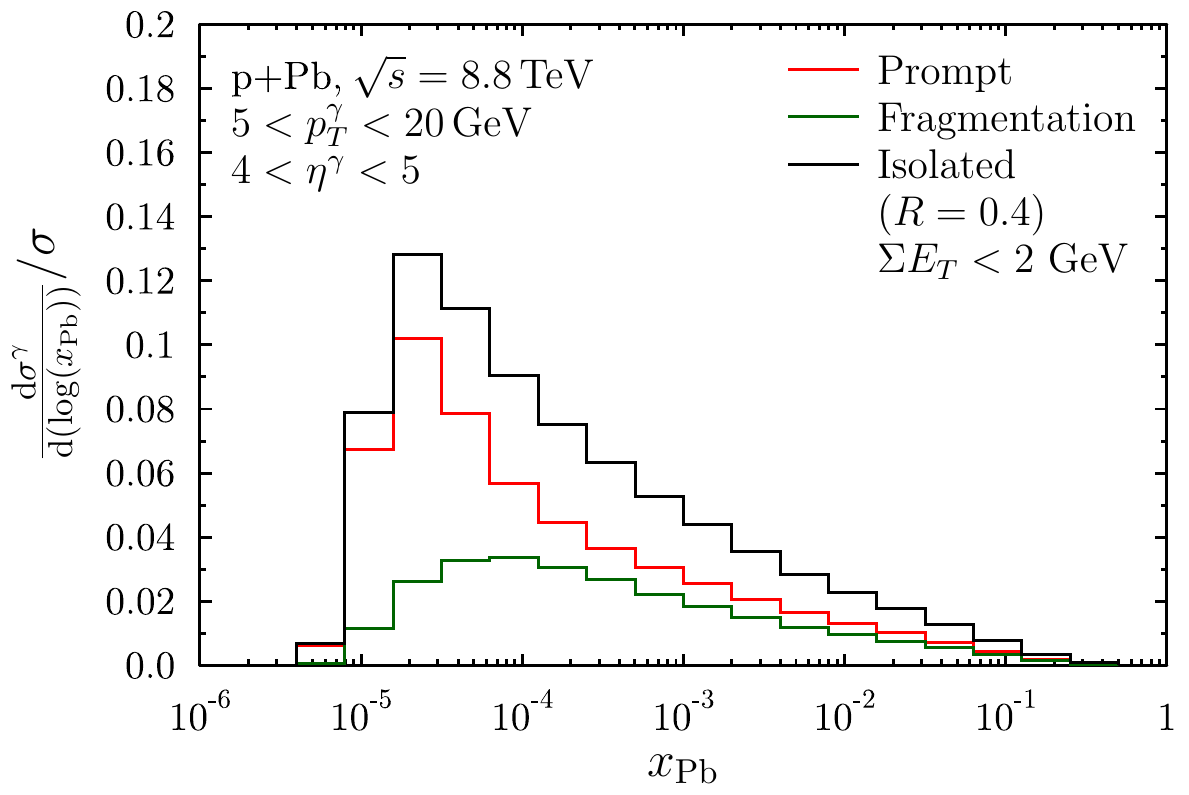}
\caption{The relative contribution to direct photon cross section from different $x_{\rm Pb}$ values for prompt (red) and fragmentation (green) component and the sum of these two (black) in p+Pb collisions at $\sqrt{s_{\rm_{NN}}}=8.8\,\mathrm{TeV}$ for inclusive (left panel) and isolated (right panel) photons. For the inclusive photons the corresponding result for $\pi^0$ production is also shown for comparison (blue dashed). From Ref.~\cite{Helenius:2014qla}.}
\label{fig:dsigmadx2Phot}
\end{center}
\end{figure}

As the dijet measurements are not very sensitive to small values of $x$ and so far there are no measurements for the hadron production in forward rapidities, the small-$x$ gluons remain still practically unconstrained. As the cross section for inclusive hadron production gets contributions also from the mid/large-$x$ region even at very forward rapidities (see figure \ref{fig:dsigmadx2pi0}) it seems not an optimal observable for the nPDF analyses. A better alternative would be direct photons for which there are two production mechanisms: prompt photons that are produced in the hard partonic scattering and fragmentation photons that are formed from the outgoing high-energy partons. The presence of the prompt component increases the relative sensitivity to the small-$x$ physics when compared to hadrons. This is demonstrated in the left panel of figure \ref{fig:dsigmadx2Phot} showing the contribution from different values of $x_{\rm Pb}$ to the differential cross section for the two components. The prompt component is indeed very sensitive to small values of $x_{\rm Pb}$ but the when the fragmentation component is added the total cross section has some contribution also from larger values of $x_{\rm Pb}$.

The contribution from the fragmentation photons can be suppressed by introducing an isolation criterion. This rejects the photons that have more than a certain amount of hadronic energy inside a cone around the photon in ($\eta$,$\phi$)-space. The fragmentation photons are typically accompanied by hadrons and the isolation condition consequently cuts part of these away. The effect on the $x_{\rm Pb}$ distributions is shown in the right panel of figure \ref{fig:dsigmadx2Phot}: the fragmentation component is suppressed more than by a factor of 2, thus increasing the sensitivity of the direct photons to small values of $x_{\rm Pb}$.

\begin{figure}[htb]
\begin{center}
\includegraphics[width=\textwidth]{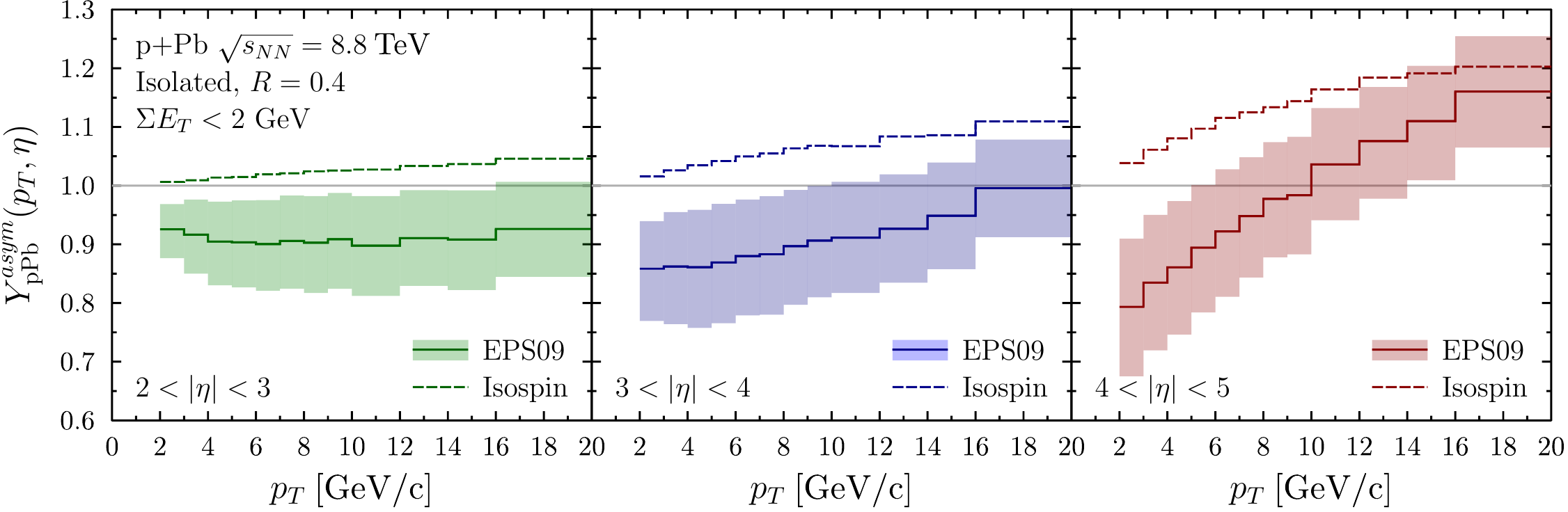}
\caption{The yield asymmetry between forward and backward rapidities in p+Pb collisions at $\sqrt{s_{\rm_{NN}}}=8.8\,\mathrm{TeV}$ for three different rapidities, $2<|\eta|<3$ (left), $3<|\eta|<4$ (middle) and $4<|\eta|<5$ (right). The isospin effect is plotted with dashed line to each panel. From Ref.~\cite{Helenius:2014qla}.}
\label{fig:yasym}
\end{center}
\end{figure}

The nuclear modification factor requires a p+p baseline which might not be available at a given energy, and if there are no luminosity measurement performed Glauber modeling for the normalization is required. These experimental difficulties can be avoided by considering the yield asymmetry between forward and backward rapidities, defined as
\begin{equation}
Y_{\rm pPb}^{\rm asym}(p_T,\eta) \equiv \left.\frac{\mathrm{d}^2\sigma_{\rm pPb}}{\mathrm{d}p_T \mathrm{d}\eta}\right|_{\eta\in[\eta_1,\eta_2]}\bigg/ \left.\frac{\mathrm{d}^2\sigma_{\rm pPb}}{\mathrm{d}p_T \mathrm{d}\eta}\right|_{\eta\in[-\eta_2,-\eta_1]},
\end{equation}
instead of $R_{\rm pPb}$. Now the photon production in backward rapidities, where the nuclear modifications are in better control from the previous measurements, serve as a baseline for the nuclear modifications at forward rapidities which in turn are sensitive to less-known small-$x$ regions. The EPS09-based predictions for this observable are presented in figure \ref{fig:yasym} for three rapidity intervals, $2<|\eta|<3$, $3<|\eta|<4$ and $4<|\eta|<5$, see Ref.~\cite{Helenius:2014qla} for details. Also the isospin effect arising from the different charge densities in proton and nuclei is shown.

\section{Conclusions}

New data sets have become available from the p+Pb run at the LHC that in principle should improve the current nPDF fits. The inclusive charged hadron $R_{\rm pPb}$ from CMS and ATLAS show, however, a surprising rise at high-$p_{\rm T}$ which is clearly beyond expectations and need to be understood before the data can be included to analyses. In general it is hard to probe different nuclear effects using the inclusive hadron production data as the cross sections are sensitive to a broad $x$-range and different effects can cancel each other out. A more promising new observable is the dijet production measured by CMS. These data can be used to reduce the uncertainty of the gluon nuclear modifications at $x>0.01$. As the direct photons at forward rapidities are more sensitive to small-$x$ gluons than inclusive hadrons, precise measurements of these would be very useful to constrain the nuclear gluon shadowing.

\section*{Acknowledgments}
I.H. is supported by the MCnetITN FP7 Marie Curie Initial Training Network, contract PITN-
GA-2012-315877.

\end{document}